\documentstyle[preprint,prd,aps]{revtex}
\begin{document}
\thispagestyle{empty}
\draft
\hoffset=-10pt
\preprint{MSUHEP-41012}
\title{Higgs-photon associated production at $e\bar{e}$ colliders}

\author{Ali Abbasabadi$^1$, David Bowser-Chao$^2$, Duane A. Dicus$^3$ and \\
Wayne W. Repko$^2$}
\address{$^1$Department of Physical Sciences \\
Ferris State University, Big Rapids, Michigan 49307}
\address{$^2$Department of Physics and Astronomy \\
Michigan State University, East Lansing, Michigan 48824}
\address{$^3$Center for Particle Physics and Department of Physics\\
University of Texas, Austin, Texas 78712}

\date{\today}
\maketitle
\begin{abstract}
We present complete analytical expressions for the amplitudes of the process
$e\bar{e}\rightarrow H\gamma$. The calculation is performed using nonlinear
gauges, which significantly simplifies both the actual analytical calculation
and the check of its gauge invariance. After comparing our results with a
previous numerical calculation, we extend the range of Higgs masses and center
of mass energies to those appropriate to LEP 200 and a future linear collider.
\end{abstract}
\pacs{13.85.Qk, 14.80.Er, 14.80.Gt}

\section{Introduction}

    The search for the Higgs-boson ($H$) at $e\bar{e}$ colliders usually
focuses
on the channel $e\bar{e}\rightarrow Z\rightarrow Z^{*}H$ at the $Z$ pole
\cite{ggpg}. Some years ago, Barroso, Pulido and Rom\~ao \cite{bpr} pointed out
that the
channel $e\bar{e}\rightarrow H\gamma$ could be important in Higgs searches, if
for no other reason than that the range of accessible $m_H$ is larger. These
authors presented numerical results for relatively low Higgs masses
($m_H\leq 60\,$ GeV) and a top quark mass $m_t$ of 40 GeV.
They concluded the signal was observable and also drew
attention to some interesting theoretical aspects of calculation related to
gauge invariance with respect to the photon field.

    Given the prospect of LEP 200 in the near term, the discussion of a higher
energy $e\bar{e}$ collider (NLC) and an indication that $m_t\sim 175\;$ GeV
\cite{top}, it seemed appropriate to revisit this calculation
with the intention of extending the range of $m_H$ and the center of mass
energy
$\sqrt{s}$\,. Accordingly, we have
computed the amplitudes for $e\bar{e}\rightarrow H\gamma$ using two nonlinear
gauges \cite{gauge,dk} for the $W$'s. Our motivation for choosing nonlinear
gauges was the reduction in the number of diagrams due to the elimination of
$W$-neutral gauge boson-charged Goldstone boson couplings. We discovered that
these gauges have the additional advantage of greatly simplifying the photon
field gauge structure of the various classes of diagrams.

    In the next Section, the choice of gauge fixing terms is presented and we
outline the strategy of our calculation. We then present numerical results for
a
range of $m_H$ and $e\bar{e}$ center of mass energies. The complete analytical
expressions for the amplitudes are given in the Appendix.

\section{Nonlinear Gauge Fixing and Calculational Approach}

    In the Standard Model, the lowest order corrections to the amplitudes for
$e\bar{e}\rightarrow H\gamma$ come from one-loop diagrams containing various
combinations of fermions, gauge bosons, Goldstone bosons ($G^{\pm},
G^0$) and ghosts ($\theta^{\pm},\bar{\theta}^{\pm}, \eta_{\gamma}, \eta_Z$).
Since the Born diagrams for this process are proportional to the electron mass
$m_e$, the one-loop corrections actually determine the amplitude in the
$m_e\rightarrow 0$ limit. For the purposes of this calculation we generally set
$m_e = 0$, although, as discussed below, some care must be exercised when doing
so in one of the diagrams.

    The precise number of diagrams encountered when computing the one-loop
corrections is influenced by the choice of the gauge for the $W^{\pm}$ and
$Z^0$. It has been shown that a carefully chosen nonlinear gauge \cite{gauge}
can
eliminate the mixed $W$-$G$-$\gamma$ vertices, reduce the number of diagrams
and
simplify the Feynman rules. More recently, a related nonlinear gauge has been
introduced which also eliminates the $W$-$G$-$Z$ vertices \cite{dk} and thereby
further reduces the number of diagrams.

    In this paper, we utilize nonlinear $R_{\xi}$ gauges specified by gauge
fixing terms of the form
\begin{equation}
{\cal L}_{GF} = -\frac{1}{2\xi_A}(f_A)^2 - \frac{1}{2\xi_Z}(f_Z)^2 -
\frac{1}{\xi_W}ff^\dagger\;,
\end{equation}
where, for either choice of nonlinear gauge,
\begin{eqnarray}
f_A & = &\; \partial_{\mu}A_{\mu} \;, \\
f_Z & = &\; \partial_{\mu}Z_{\mu} + \xi_Zm_ZG^0\;.
\end{eqnarray}
For the nonlinear gauges, we choose
\begin{equation}
f =\; \partial_{\mu}W_{\mu} - igW_{\mu}^3W_{\mu} + i\xi_Wm_WG\;,
\end{equation}
or
\begin{equation}
f =\; \partial_{\mu}W_{\mu} - ig^\prime B_{\mu}W_{\mu} + i\xi_Wm_WG\;,
\end{equation}
with
\begin{eqnarray}
W_{\mu}^3 & = &\; \cos\theta_W Z_{\mu} + \sin\theta_W A_{\mu}\;,\\
B_{\mu}   & = &\; -\sin\theta_W Z_{\mu} + \cos\theta_W A_{\mu}\;,
\end{eqnarray}
and $\theta_W$ denoting the weak mixing angle.
The gauge parameters $\xi_A, \xi_Z$ and $\xi_W$ are all chosen to be unity,
which corresponds to nonlinear 't Hooft-Feynman gauges. We obtain the ghost
couplings for a particular nonlinear gauge by examining the revelant gauge
fixing terms.

    The diagrams encountered in the calculation of the $e\bar{e}\rightarrow
H\gamma$ amplitudes fall into three catagories (illustrated in Fig. 1): those
with two poles at $m_Z$;
those with a single $Z^0$ or $\gamma$ pole; and triangle or box diagrams with
gauge bosons and fermions in the loops. Apart from a reduction in the number
of diagrams to
be calculated, the use of nonlinear gauges has the additional feature that
these catagories are {\em separately} gauge invariant with respect to the
photon
field. This is not the case in linear gauges, where the replacement of the
photon polarization vector $\hat{\epsilon}_{\mu}(k)$ by the momentum vector
$k_{\mu}$ mixes contributions from various catagories to produce the expected
zero result.

    It is worth mentioning that the
separation of the diagrams into gauge invariant subsets can be traced to
the simplified Ward identity satisfied by the $W$-$W$-$\gamma$ vertex
in either nonlinear gauge. Specifically, if $\Gamma_{\mu\nu\lambda}(p + k, p)$
denotes the vertex function for an incoming $W$ with momentum $p$ and
polarization $\hat{\epsilon}_{\nu}(p)$, an outgoing $W$ of momentum $p + k$ and
polarization $\hat{\epsilon}_{\mu}(p + k)$ and a photon of momentum $k$ and
polarization $\hat{\epsilon}_{\lambda}(k)$, the Ward identity is
\begin{equation} \label{ward}
k_{\lambda}\Gamma_{\mu\nu\lambda}(p + k, p) =\; e\delta_{\mu\nu}
[\Delta_F^{-1}(p + k) - \Delta_F^{-1}(p)]\;,
\end{equation}
where
\begin{equation}
\Delta_F^{-1}(p) =\; p^2 + m_W^2\;.
\end{equation}
The simplicity of the righthand side of Eq.(\ref{ward}) means that the
replacement of the photon polarization vector $\hat{\epsilon}(k)$ by the
momentum $k$ merely eliminates one or the other of the propagators adjacent to
the vertex. There are no extra contributions which serve to connect one
catagory
of diagrams to another.

\subsection{Double pole diagrams}

    One consequence of the separation mentioned above is that the double
pole diagrams illustrated in Fig. 1\,(a) vanish when the photon is on the mass
shell. There is no need to perform a renormalization, unlike the situation
found in Ref. \cite{bpr}.

    On the photon mass shell, any given double pole diagram makes
a contribution of the general form
\begin{equation}
{\cal M}_{\mu\nu} = (eg^2/m_Z)(\alpha + \beta\tan^2\theta_W)\delta_{\mu\nu}.
\end{equation}
The coefficients $\alpha$ and $\beta$ are expressible in terms of the scalar
loop integrals $A_0(m_W^2) = A_0$ and $B_0(k^2 = 0,m_W^2,m_W^2) = B_0$ of
't Hooft and Veltman \cite{hv}. Although $A_0$ and $B_0$ depend on the number
of
dimensions $n$, they satisfy the relation
\begin{equation} \label{van}
(1 - \frac{n}{2})A_0 - m_W^2B_0 = 0\;.
\end{equation}
The contributions of the non-vanishing diagrams are listed in Table I. These
consist of tadpole and bubble diagrams with $W$'s, charged Goldstone bosons or
ghosts in the loops. Notice that for each particle the sum of the
one-point and two-point contributions results in the combination
Eq. (\ref{van})\,. Consequently, the double pole diagrams all vanish.

\subsection{Single pole diagrams}

    The single pole diagrams of Fig. 1\,(b) involve $Z^0$ or $\gamma$ poles,
with the virtual gauge boson of momentum $p$ decaying into $H\,\gamma$ (momenta
$k^{\prime}$ and $k$). Our result for the $\gamma^*\rightarrow H\gamma$
amplitude is gauge invariant, in agreement with Ref.\cite{bpr}.
Unlike Ref. \cite{bpr}, we also find the $Z^*\rightarrow H\gamma$
amplitude to be gauge invariant. In the nonlinear gauges there is no need to
combine this contribution with box contributions to achieve a gauge invariant
result.

    The relevant diagrams consist of quark, $W$, Goldstone boson and ghost
loops, which we evaluate using the Passarino and Veltman decomposition
\cite{pv}. The complete contribution from the photon pole is
\begin{eqnarray}\label{gpole}
{\cal M}_{\rm Pole}^{\gamma} & = &\; \frac{\alpha^2m_W}{\sin\theta_W}
\bar{v}(p_2)\gamma_{\mu}u(p_1)\hat{\epsilon}_{\nu}(k)\Bigl({\delta_{\mu\nu}
k\!\cdot\!(p_1 + p_2) - k_{\mu}(p_1 + p_2)_{\nu}\over s}\Bigr) \\ \nonumber
&  &\; \times\Bigl\{4(6 + \frac{m_H^2}{m_W^2})C_{23}(s,m_H^2,m_W^2) -
16C_0(s,m_H^2,m_W^2) \\ \nonumber
&  &\;-\,\frac{8}{3}\frac{m_t^2}{m_W^2}\Bigl(4C_{23}(s,m_H^2,m_t^2) -
C_0(s,m_H^2,m_t^2)\Bigr)\Bigr\}\;,
\end{eqnarray}
where $s = -(p_1 + p_2)^2$ and $C_0(s,m_H^2,m^2)$ and $C_{23}(s,m_H^2,m^2)$
are scalar functions given in the Appendix A.

    For the $Z^0$ pole contribution, we find
\begin{eqnarray}\label{zpole}
{\cal M}_{\rm Pole}^{Z} & = &\; \frac{\alpha^2m_W}{\sin^3\theta_W}\bar{v}(p_2)
\gamma_{\mu}(v_e + \gamma_5)u(p_1)\hat{\epsilon}_{\nu}(k)
\Bigl({\delta_{\mu\nu}k\!\cdot\!(p_1 + p_2) - k_{\mu}(p_1 + p_2)_{\nu}\over
s - m_Z^2 + im_Z\Gamma_Z}\Bigr) \\ \nonumber
&  &\; \times\Bigl\{\Bigl((6 - \frac{1}{\cos^2\theta_W})
+ \frac{m_H^2}{2m_W^2}\frac{1 - 2\sin^2\theta_W}{\cos^2\theta_W}\Bigr)
C_{23}(s,m_H^2,m_W^2) \\ \nonumber
&  &\;+\, \Bigl(\frac{1}{\cos^2\theta_W} - 4\Bigr)
C_0(s,m_H^2,m_W^2) \\ \nonumber
&  &\;-\,\frac{1}{4}\frac{m^2_t}{m^2_W}\frac{1 - (8/3)\sin^2\theta_W}
{\cos^2\theta_W}\Bigl(4C_{23}(s,m_H^2,m_t^2) - C_0(s,m_H^2,m_t^2)\Bigr)\Bigr\}
\;.
\end{eqnarray}
Here, $v_e = 1 - 4\sin^2\theta_W$ and $\Gamma_Z$ denotes the $Z^0$ width.

\subsection{Box and associated triangle diagrams}

    The remaining diagrams (Fig. 1\,(c)) have no gauge boson poles. They
consist of
box diagrams where the photon emerges from one of the box vertices together
with associated triangle diagrams with the photon being radiated by one of the
incoming leptons. There are two such combinations of boxes and triangles: one
with $Z$'s in the loops and one with $W$'s in the loops.

    The contribution to the matrix element from the box and triangles with
internal $Z$'s is given by
\begin{eqnarray} \label{zbox}
{\cal M}_{\rm Box}^Z & =
&\;-\,\frac{\alpha^2m_Z}{4\sin^3\theta_W\cos^3\theta_W}
\bar{v}(p_2)\gamma_{\mu}(v_e + \gamma_5)^2u(p_1)\hat{\epsilon}_{\nu}(k)
\\ \nonumber
&   &\;\times\,\Bigl\{\Bigl(\delta_{\mu\nu}k\!\cdot\!p_1 - k_{\mu}(p_1)_{\nu}
\Bigr)A(s,t,u) + \Bigl(\delta_{\mu\nu}k\!\cdot\!p_2 - k_{\mu}(p_2)_{\nu}
\Bigr)A(s,u,t)\Bigr\}\;,
\end{eqnarray}
with $t = -(p_1 - k)^2$ and $u = -(p_2 - k)^2$.
The scalar function $A(s,t,u)$ is given in the Appendix B. In
the evaluation of the loop integrals for this contribution, we encountered
$\ln(m_e^2)$ terms in the box diagram. These were shown to cancel
in the final result and we then set $m_e = 0$ in the remaining expressions.
Notice, too, that Eq. (\ref{zbox}) is explicitly gauge invariant.

    The contribution from the diagrams with internal $W$'s is also gauge
invariant and given by
\begin{eqnarray} \label{wbox}
{\cal M}_{\rm Box}^{W} & = &\;\frac{\alpha^2m_W}{2\sin^3\theta_W}\bar{v}(p_2)
\gamma_{\mu}(1 + \gamma_5)^2u(p_1)\hat{\epsilon}_{\nu}(k) \\ \nonumber
&   &\;\times\,\Bigl\{\Bigl(\delta_{\mu\nu}k\!\cdot\!p_1 - k_{\mu}(p_1)_{\nu}
\Bigr)\Bigl(A_1(s,t,u) + A_2(s,u,t)\Bigr) \\ \nonumber
&   &\;+\,\Bigl(\delta_{\mu\nu}k\!\cdot\!p_2 - k_{\mu}(p_2)_{\nu}\Bigr)\Bigl(
A_2(s,t,u) + A_1(s,u,t)\Bigr)\Bigr\}\;,
\end{eqnarray}
with expressions for $A_1(s,t,u)$ and $A_2(s,t,u)$ given in
the Appendix C.

    All expressions for the scalar functions given in the Appendices have been
checked numerically using Veltman's program FORMF \cite{loop} and Vermaseren's
program FF \cite{ff}. The results presented below were obtained using the
formulae in Appendices A-C.

\section{Discussion}

    The differential cross section $d\sigma(e\bar{e}\rightarrow H\gamma)
/d\Omega_{\gamma}$ is given by
\begin{equation}\label{dsigma}
\frac{d\sigma(e\bar{e}\rightarrow H\gamma)}{d\Omega_{\gamma}} =\;
\frac{1}{256\pi^2}\frac{s - m_H^2}{s^2}\,\sum_{\rm spin}|{\cal M}|^2\;,
\end{equation}
where the invariant amplitude ${\cal M}$ is the sum of Eqs.
(\ref{gpole}-\ref{wbox}). Explicitly, we have
\begin{eqnarray}
\sum_{\rm spin}|{\cal M}|^2 & = &\;\frac{\alpha^4\,m_W^2\,s}{16\sin^6\theta_W\,
\cos^8\theta_W}\biggl\{(t^2 + u^2)\biggl(|{\cal A}_{\gamma}|^2 + 2v_e{\rm Re}
({\cal A}_{\gamma}{\cal A}_Z^*) + (1 + v_e^2)|{\cal A}_Z|^2\biggr) \\ \nonumber
&  &\;+ t^2\biggl[2(1 + v_e^2){\rm Re}({\cal A}_{\gamma}{\cal A}^*)+ 4{\rm Re}
({\cal A}_{\gamma}{\cal A}_{12}^*) + 2(v_e^3 + 3v_e){\rm Re}({\cal A}_Z
{\cal A}^*)  \\ \nonumber
&  &\;+ 4(1 + v_e){\rm Re}({\cal A}_Z{\cal A}_{12}^*)
+ (1 + 6v_e^2 + v_e^4)|{\cal A}|^2 + 4(1 + v_e)^2{\rm Re}({\cal A}{\cal
A}_{12}^*) \\ \nonumber
&  &\;+ 8|{\cal A}_{12}|^2\biggr] + u^2\biggl[t\leftrightarrow u\biggr]
\biggr\}\;.
\end{eqnarray}
Here,  ${\cal A}_{\gamma}$, ${\cal A}_Z$, ${\cal A}$ and ${\cal A}_{12}$ are
\begin{eqnarray}
{\cal A}_{\gamma} & = &\;4\sin^2\theta_W\,\cos^4\theta_W\frac{1}{s}\times
\mbox{\{$\cdots$\}}\;, \label{agam} \\
{\cal A}_Z & = &\;4\cos^4\theta_W\frac{1}{s - m^2_Z + im_Z\Gamma_Z}\times
\mbox{\{$\cdots$\}}\;, \label{azee} \\
{\cal A} & = &\;A(s,t,u)\;,\quad\mbox{(Eq.(\ref{astu1}))}\;, \\
{\cal A}_{12} & = &\; -2\cos^4\theta_W\left(A_1(s,t,u) + A_2(s,u,t)\right)
\quad\mbox{(Eqs.(\ref{a1}) \& (\ref{a2}))}\;,
\end{eqnarray}
and \{$\cdots$\} in Eqs.~(\ref{agam}) and (\ref{azee}) denotes the contents of
the curly brackets in Eqs.~(\ref{gpole}) and (\ref{zpole}), respectively.

    For purposes of comparison with Table 4 of Ref. \cite{bpr}, Eq.
(\ref{dsigma}) was used to compute $\sigma(e\bar{e}\rightarrow
H\gamma)$ when $m_H = m_t = 40\;$ GeV and $60\; \mbox{GeV} \leq \sqrt{s} \leq
150\;\mbox{GeV}$\,. The two sets of results are given in Table II. From this
data, it is evident that there are differences between the two calculations on
a
contribution-by-contribution basis. We attribute this variation to the fact
that
the ${\cal M}^Z_{\rm Pole}$ and ${\cal M}^Z_{\rm Box}$ are not separately gauge
invariant for the linear 't Hooft-Feynman gauge of Ref. \cite{bpr}. These
amplitudes are separately gauge invariant for the nonlinear 't Hooft-Feynman
gauges used in the present calculation. The total cross section for the two
cases is plotted in Fig. 2. The results are in good agreement, although the
cross section of Ref. \cite{bpr} is slightly lower than ours both below and
above the $Z$ peak.

    In Fig. 3, the cross sections corresponding to  $m_H = 60, 80, 100, 120\;
\mbox{and}\;150\;
\mbox{GeV}$ are plotted for the LEP 200 range $\sqrt{s}\leq 200\,$GeV using a
top quark mass of 174 GeV. Apart from $m_H = 150\;$ GeV, the cross sections
rise
to a level greater than 0.1 fb as $\sqrt{s}\rightarrow 200\;$ GeV. The only
discernible feature above the $Z$ peak is the 2$W$ threshold at $\sqrt{s}\sim
160\;\mbox{GeV}$\,.

    The extension to the range $\sqrt{s}\leq 500\,$ GeV is shown in Fig. 4. All
curves clearly show the effect of the top threshold. In addition, the $m_H = 2
M_Z$ curve forms a new upper envelope for the cross sections corresponding to
the region $m_H\geq 2m_Z$.

    The cross sections are, of course, quite small. At expected LEP 200
luminosities (0.5 - 1.0 fb$^{-1}$) no events of this type should be seen. Thus,
any observed $H\,\gamma$ events would be a signal of new physics. For proposed
NLC luminosities (50 fb$^{-1}$/yr), $H\,\gamma$ events should be seen. At other
energies or masses, our analytic results allow the cross section to be
determined easily.

\acknowledgments

We are greatful to Dr. J. Pulido and Dr. J. C. Rom\~ao for supplying us with
details of their calculation and Dr. C.-P. Yuan for numerous helpful
discussions. One of us (A. A.) wishes to thank the Department
of Physics and Astronomy at Michigan State University for its support and
hospitality. This work was supported in part by the National
Science Foundation under grant PHY-93-07980, by the United States Department
of Energy under contract DE-FG013-93ER40757 and by the Texas National Research
Laboratory Commission under grant RGFY93-331.

\appendix\section{Single pole diagram integrals}

    When evaluating the gauge boson pole contributions, we encounter the
functions $C_0(s,m_H^2,m^2)$ and $C_{23}(s,m_H^2,m^2)$. These functions are
defined in terms of loop integrals as
\begin{eqnarray}
C_0(s,m_H^2,m^2) & = &\;\frac{1}{i\pi^2}\int d^{\,n}q{1\over (q^2 + m^2)
((q - p)^2 + m^2)((q - k)^2 + m^2)}\;, \\
C_{\mu\nu}(p,k) & = &\;\frac{1}{i\pi^2}\int d^{\,n}q{q_{\mu}q_{\nu}
\over (q^2 + m^2)((q - p)^2 + m^2)((q - k)^2 + m^2)} \\ \nonumber
& = &\;C_{21}(s,m_H^2,m^2)p_{\mu}p_{\nu} + C_{22}(s,m_H^2,m^2)k_{\mu}k_{\nu} +
 \\ \nonumber
&   &\;C_{23}(s,m_H^2,m^2)(p_{\mu}k_{\nu} + k_{\mu}p_{\nu}) +
C_{24}(s,m_H^2,m^2)\delta_{\mu\nu} \;,
\end{eqnarray}
where we have used $p^2 = -s$ and the mass shell conditions $k^{\prime\,2} =
-m_H^2$, $k^2 = 0$. The latter enables us to
eliminate reference to the $k^2$ dependence of the $C$'s. It should be noted
that numerous other $C$'s appear in the course of the calculation, but only
$C_0$ and $C_{23}$ survive in the final result.

    The evaluation of $C_0(s,m_H^2,m^2)$ is straightforward, yielding
\begin{equation}
C_0(s,m_H^2,m^2)  =\;\frac{1}{(s - m_H^2)}\Bigl(C(\frac{m_H^2}{m^2}) -
C(\frac{s}{m^2})\Bigr)\;,
\end{equation}
where
\begin{eqnarray}
C(\beta) & = &\;\int_0^1\frac{dx}{x}\ln\Bigl(1 - \beta x(1 - x) -
i\varepsilon\Bigr) \\
         & = &\;\left\{
\begin{array}{lll}
-2\Bigl(\sin^{-1}(\sqrt{\frac{\displaystyle\beta}{\displaystyle 4}}\,)\Bigr)^2
 &  & 0\leq\beta\leq 4 \\
2\Bigl(\cosh^{-1}(\sqrt{\frac{\displaystyle\beta}{\displaystyle 4}}\,)\Bigr)^2
- \frac{\displaystyle\pi^2}{\displaystyle2} - 2i\pi\cosh^{-1}(\sqrt{
\frac{\displaystyle\beta}{\displaystyle 4}}) &  & \beta\geq 4
\end{array}
\right.\,.
\end{eqnarray}
By analyzing Eq.(A2), it can be shown that $C_{23}(s,m_H^2,m^2)$ is
expressible as
\begin{eqnarray}
C_{23}(s,m_H^2,m^2) & = &\;-\frac{1}{2(s - m_H^2)}\Bigl[-(2 - \frac{n}{2})
B_0(k^{\prime\;2},m^2) + 2m^2C_0(s,m_H^2,m^2)\\ \nonumber
                    &   &\;+ \frac{s}{(s - m_H^2)}\Bigl(B_0(k^{\prime\;2},m^2)
- B_0(p^2,m^2)\Bigr)\Bigr]\;.
\end{eqnarray}
The two-point scalar function $B_0(p^2,m^2)$ is defined by
\begin{equation}
B_0(p^2,m^2)  =\;\frac{1}{i\pi^2}\int d^{\,n}q{1\over (q^2 + m^2)
((q + p)^2 + m^2)}\;.
\end{equation}
This integral can be evaluated to give
\begin{equation}
B_0(p^2,m^2)  =\;\frac{\pi^{(n/2 - 2)}\Gamma(2 - \frac{\displaystyle n}
{\displaystyle 2})}{(m^2)^{(2 - n/2)}} - B(-\frac{p^2}{m^2})\;,
\end{equation}
with
\begin{eqnarray}
B(\beta) & = & \int_0^1dx\ln\bigl(1 - \beta x(1 - x) - i\varepsilon\Bigr) \\
         & = & \left\{
\begin{array}{lll}
2\Bigl[\sqrt{\frac{\displaystyle 4 - \beta}{\displaystyle\beta}}\sin^{-1}
(\sqrt{\frac{\displaystyle\beta}{\displaystyle 4}}\,) - 1\Bigr] &  &
0\leq\beta\leq 4 \\
2\Bigl[\sqrt{\frac{\displaystyle\beta - 4}{\displaystyle\beta}}\cosh^{-1}
(\sqrt{\frac{\displaystyle\beta}{\displaystyle 4}}\,) - 1 -
\frac{\displaystyle i\pi}{\displaystyle 2}\sqrt{\frac{\displaystyle\beta - 4}
{\displaystyle\beta}}\,\Bigr] &  & \beta\geq 4
\end{array}
\right.\,.
\end{eqnarray}
$C_{23}(s,m_H^2,m^2)$ then takes the form
\begin{eqnarray}
C_{23}(s,m_H^2,m^2) & = &\;\frac{1}{2}\frac{1}{(s - m_H^2)}\Bigl[1 +
\frac{s}{(s
- m_H^2)}\Bigl(B(\frac{m_H^2}{m^2}) - B(\frac{s}{m^2})\Bigr) \\ \nonumber
&   &\;-\frac{2m^2}{(s - m_H^2)}\Bigl(C(\frac{m_H^2}{m^2}) - C(\frac{s}{m^2})
\Bigr)\Bigr]\;.
\end{eqnarray}

\section{Z Box}

    The diagrams with $Z$'s in the loop consist of the crossed box illustrated
in the first of Figs.\,1\,(c) and two triangle diagrams of the type shown in
the last
of Figs.\,1\,(c). In the nonlinear gauges, the triangle contributions serve to
cancel a non-gauge-invariant term in the crossed box. The entire contribution
is given by the function $A(s,t,u)$, which appears in Eq. (\ref{zbox}). In
terms
of the decomposition of Ref. \cite{pv}, this function is given by
\begin{eqnarray}\label{astu}
A(s,t,u) & = &\;D_0(s,t,u,m_H^2,m_e^2,m_Z^2) + D_{11}(s,t,u,m_H^2,m_e^2,m_Z^2)
\\ \nonumber
&   &+ D_{12}(s,t,u,m_H^2,m_e^2,m_Z^2) + D_{24}(s,t,u,m_H^2,m_e^2,m_Z^2)\;.
\end{eqnarray}
When the $D_{\alpha\beta}$ in Eq.\,(\ref{astu}) are expanded in terms of scalar
integrals, the expression for $A(s,t,u)$ takes the form
\begin{eqnarray}\label{astu1}
A(s,t,u) & = &\;\frac{1}{2}\frac{1}{st}\Bigl[\frac{1}{t}(t - m_Z^2)
\Bigl(m_Z^2(t + u) - tu\Bigr)D_0(1,2,3,4) - (t - m_Z^2)C_0(1,2,3) \\ \nonumber
&  &\;-\frac{1}{t}\Bigl((t - m_Z^2)(s - t) - 2m_Z^2\frac{st}{(s + t)}\Bigr)
C_0(1,2,4) + \frac{1}{t}(t - m_Z^2)(s + u)C_0(1,3,4) \\ \nonumber
&  &\;-\frac{u}{t}(t - m_Z^2)C_0(2,3,4)
+ \frac{2s}{(s + t)}\Bigl(B_0(1,4) - B_0(2,4)\Bigr)\Bigr]\;,
\end{eqnarray}
where we have used the compact notation of Ref.\,\cite{pv} with $m_1 = m_4 =
m_Z$ and $m_2 = m_3 = m_e$. This labeling is illustrated in Fig. 5\,(a).

    Explicit evaluation of $D_0(1,2,3,4)$ gives
\begin{eqnarray} \label{d0z}
D_0(1,2,3,4) & = &\;\frac{1}{(tu - m_Z^2t - m_Z^2u)}\Biggl\{\Biggl[\ln\left(
1 - \frac{t}{m_Z^2}\right)
\left(\ln\Bigl(\frac{tu}{m_e^2m_H^2}\Bigr) +
2\ln\Bigl(1 - \frac{m_Z^2}{t}\Bigr)\right.\nonumber \\
&  &\;\left.+ \ln\Bigl(-\frac{m_Z^2}{t}\Bigr) + \ln\Bigl(1 - \frac{m_Z^2}{t} -
\frac{m_Z^2}{u}\Bigr) - \ln\Bigl(\beta_+ - \frac{m_Z^2}{t}\Bigr) -
\ln\Bigl(\beta_- - \frac{m_Z^2}{t}\Bigr)\right)
\nonumber \\
&  &\;+ 2Li_2\left(\frac{\mbox{\rule[-5pt]{0pt}{14pt}} - \frac{\displaystyle
\mbox{\rule[-5pt]{0pt}{14pt}}m_Z^2}{\displaystyle\mbox{\rule{0pt}{9pt}}t}}
{\mbox{\rule{0pt}{9pt}}1 - \frac{\displaystyle\mbox{\rule[-5pt]{0pt}{14pt}}
m_Z^2}{\displaystyle\mbox{\rule{0pt}{9pt}}t}}\right)
+ Li_2\left(\frac{\mbox{\rule[-5pt]{0pt}{14pt}} - \frac{\displaystyle
\mbox{\rule[-5pt]{0pt}{14pt}}m_Z^2}{\displaystyle\mbox{\rule{0pt}{9pt}}t}}
{\mbox{\rule{0pt}{9pt}}1 - \frac{\displaystyle\mbox{\rule[-5pt]{0pt}{14pt}}
m_Z^2}{\displaystyle\mbox{\rule{0pt}{9pt}}t} - \frac{\displaystyle
\mbox{\rule[-5pt]{0pt}{14pt}}m_Z^2}{\displaystyle\mbox{\rule{0pt}{9pt}}u}}
\right)
- Li_2\left(\frac{\mbox{\rule[-5pt]{0pt}{14pt}} - \frac{\displaystyle
\mbox{\rule[-5pt]{0pt}{14pt}}m_Z^2}{\displaystyle\mbox{\rule{0pt}{9pt}}t}}
{\mbox{\rule{0pt}{9pt}}\beta_+ - \frac{\displaystyle\mbox{\rule[-5pt]{0pt}
{14pt}}m_Z^2}{\displaystyle\mbox{\rule{0pt}{9pt}}t}}\right)
\nonumber \\
&  &\;- Li_2\left(\frac{\mbox{\rule[-5pt]{0pt}{14pt}} - \frac{\displaystyle
\mbox{\rule[-5pt]{0pt}{14pt}}m_Z^2}{\displaystyle\mbox{\rule{0pt}{9pt}}t}}
{\mbox{\rule{0pt}{9pt}}\beta_- - \frac{\displaystyle\mbox{\rule[-5pt]{0pt}
{14pt}}m_Z^2}{\displaystyle\mbox{\rule{0pt}{9pt}}t}}\right)
- Li_2\left(\frac{\mbox{\rule[-5pt]{0pt}{14pt}}1 - \frac{\displaystyle
\mbox{\rule[-5pt]{0pt}{14pt}}m_Z^2}{\displaystyle\mbox{\rule{0pt}{9pt}}t}}
{\mbox{\rule{0pt}{9pt}}1 - \frac{\displaystyle\mbox{\rule[-5pt]{0pt}{14pt}}
m_Z^2}{\displaystyle\mbox{\rule{0pt}{9pt}}t} - \frac{\displaystyle
\mbox{\rule[-5pt]{0pt}{14pt}}m_Z^2}{\displaystyle\mbox{\rule{0pt}{9pt}}u}}
\right)\nonumber \\
&  &\;
+ Li_2\left(\frac{\mbox{\rule[-5pt]{0pt}{14pt}}1 - \frac{\displaystyle
\mbox{\rule[-5pt]{0pt}{14pt}}m_Z^2}{\displaystyle\mbox{\rule{0pt}{9pt}}t}}
{\mbox{\rule{0pt}{9pt}}\beta_+ - \frac{\displaystyle\mbox{\rule[-5pt]{0pt}
{14pt}}m_Z^2}{\displaystyle\mbox{\rule{0pt}{9pt}}t} + i\varepsilon}\right)
+ Li_2\left(\frac{\mbox{\rule[-5pt]{0pt}{14pt}}1 - \frac{\displaystyle
\mbox{\rule[-5pt]{0pt}{14pt}}m_Z^2}{\displaystyle\mbox{\rule{0pt}{9pt}}t}}
{\mbox{\rule{0pt}{9pt}}\beta_- - \frac{\displaystyle\mbox{\rule[-5pt]{0pt}
{14pt}}m_Z^2}{\displaystyle\mbox{\rule{0pt}{9pt}}t} - i\varepsilon}\right)
- \frac{\pi^2}{3}\Biggr] + \left[t\leftrightarrow u\right]\Biggr\}\;,
\end{eqnarray}
where $Li_2(z)$ is the dilogarithm or Spence function \cite{dilog} defined by
\begin{equation}
Li_2(z) =\;-\int_0^1\!\frac{dt}{t}\ln(1 - zt)\;.
\end{equation}
Notice that $D_0(1,2,3,4)$ has a $\ln(m_e^2)$ dependence, which eventually
cancels as shown below. The roots $\beta_\pm$ are
\begin{equation} \label{zroots}
\beta_\pm  = \;\frac{1}{2}\left(1\,\pm\, \sqrt{1 - \frac{\displaystyle
\mbox{\rule[-3pt]{0pt}{12pt}}4m_Z^2}{\displaystyle\mbox{\rule{0pt}{9pt}}m_H^2}}
\;\right)\;.
\end{equation}

    The various $C_0$ functions are
\begin{eqnarray} \label{c0z1}
C_0(1,2,3) & = &\;\frac{1}{t}\Biggl[
Li_2\left(\mbox{\rule{0pt}{21pt}}\frac{\mbox{\rule[-5pt]{0pt}
{14pt}}1}{\mbox{\rule{0pt}{9pt}}1 - \frac{\displaystyle\mbox{\rule[-5pt]{0pt}
{14pt}}m_Z^2}{\displaystyle\mbox{\rule{0pt}{9pt}}t}}\right) -
\frac{1}{2}\ln^2\left(1 - \frac{t}{m_Z^2}\right) + \ln\left(1 - \frac{t}{m_Z^2}
\right)\ln\left(\frac{m_e^2}{m_Z^2}\right)\Biggr]\,, \\ [6pt]
\label{c0z3}
C_0(1,2,4) & = &\;\frac{1}{m_H^2 - u}\Biggl[-Li_2\left(\frac{\alpha_1 -
1}{\alpha_1 - i\varepsilon}\right) - Li_2\left(\frac{\alpha_2}{\alpha_2 -
\beta_- + i\varepsilon}\right) \nonumber \\ [4pt]
&  &\hspace*{.75in}+ Li_2\left(\frac{\alpha_2 - 1}{\alpha_2 - \beta_- +
i\varepsilon}\right) - Li_2\left(\frac{\alpha_2}{\alpha_2 - \beta_+ - i
\varepsilon}\right) \\ [4pt]
&  &\hspace*{.75in}+ Li_2\left(\frac{\alpha_2 - 1}{\alpha_2 - \beta_+ -
i\varepsilon}\right) + Li_2\left(\frac{\alpha_3}{\alpha_3 - 1 + i\varepsilon}
\right) \nonumber \\ [4pt]
&  &\;\hspace*{.75in} + Li_2\left(\frac{\alpha_3}{\alpha_3 -
\frac{\displaystyle\mbox{\rule[-5pt]{0pt}{14pt}}m_Z^2}{\displaystyle
\mbox{\rule{0pt}{9pt}}u} -i\varepsilon}\right) - Li_2\left(\frac{\alpha_3 - 1}
{\alpha_3 - \frac{\displaystyle\mbox{\rule[-5pt]{0pt}{14pt}}m_Z^2}
{\displaystyle\mbox{\rule{0pt}{9pt}}u} - i\varepsilon}\right)\Biggr]\,,
\nonumber \\ [6pt]
C_0(1,3,4) & = &\;\frac{1}{m_H^2 - t}\Biggl[- Li_2\left(\frac{\gamma - 1}
{\gamma - 1 + \frac{\displaystyle\mbox{\rule[-5pt]{0pt}{14pt}}m_Z^2}
{\displaystyle\mbox{\rule{0pt}{9pt}}t} + i\varepsilon}\right) +
Li_2\left(\frac{\gamma}{\gamma - 1 + \frac{\displaystyle\mbox{\rule[-5pt]{0pt}
{14pt}}m_Z^2}{\displaystyle\mbox{\rule{0pt}{9pt}}t} + i\varepsilon}\right)
\nonumber \\ [4pt]
&  &\hspace*{.75in}- Li_2\left(\frac{\gamma - 1}{\gamma - i\varepsilon}\right)
-
Li_2\left(\frac{\gamma}{\gamma - \beta_- + i\varepsilon}\right) \\ [4pt]
&  &\hspace*{.75in} + Li_2\left(\frac{\gamma - 1}{\gamma - \beta_- + i
\varepsilon}\right) - Li_2\left(\frac{\gamma}{\gamma - \beta_+ - i\varepsilon}
\right) \nonumber \\ [4pt]
&  &\hspace*{.75in} + Li_2\left(\frac{\gamma - 1}{\gamma - \beta_+ -
i\varepsilon}\right) + \frac{\pi^2}{6}\Biggr]\,, \nonumber \\ [6pt]
\label{c0z2}
C_0(2,3,4) & = &\;\frac{1}{u}\Biggl[Li_2\left(\frac{\mbox{\rule[-5pt]{0pt}
{14pt}}1}{\mbox{\rule{0pt}{9pt}}1 - \frac{\displaystyle\mbox{\rule[-5pt]{0pt}
{14pt}}m_Z^2}{\displaystyle\mbox{\rule{0pt}{9pt}}u}}\right) -
\frac{1}{2}\ln^2\left(1 - \frac{u}{m_Z^2}\right) + \ln\left(1 - \frac{u}{m_Z^2}
\right)\ln\left(\frac{m_e^2}{m_Z^2}\right)\Biggr]\,.
\end{eqnarray}
The roots $\alpha_1$, $\alpha_2$, $\alpha_3$ and $\gamma$ are
\begin{equation}
\left.
\begin{array}{lcl}
\alpha_1 = \frac{\displaystyle\mbox{\rule[-5pt]{0pt}{14pt}}u^2 - m_H^2u +
m_Z^2m_H^2}{\displaystyle\mbox{\rule{0pt}{9pt}}(m_H^2 - u)^2}\,,
& &\alpha_2 = \frac{\displaystyle\mbox{\rule[-5pt]{0pt}{14pt}}m_Z^2}
{\displaystyle\mbox{\rule{0pt}{9pt}}(m_H^2 - u)}\,, \\ [8pt]
\alpha_3 = \frac{\displaystyle\mbox{\rule[-5pt]{0pt}{14pt}}m_Z^2m_H^2}
{\displaystyle\mbox{\rule{0pt}{9pt}}u(m_H^2 - u)}\,,
& &\gamma = \frac{\displaystyle\mbox{\rule[-5pt]{0pt}{14pt}}m_Z^2}
{\displaystyle\mbox{\rule{0pt}{9pt}}(m_H^2 - t)}\,,
\end{array}\right.
\end{equation}
and the roots $\beta_{\pm}$ are those of Eq. (\ref{zroots}). Note that
$C_0(1,2,4)$ can also be obtained directly from $C_0(1,3,4)$ by replacing $t$
with $u$.

    Finally, the required $B_0$ functions are
\begin{eqnarray} \label{b0z}
B_0(1,4) & = &\;\Delta + \beta_+\ln\left(\frac{\mbox{\rule[-5pt]{0pt}{14pt}}
\beta_+ - 1 + i\varepsilon}{\mbox{\rule{0pt}{9pt}}\beta_+}\right) +
\beta_-\ln\left(\frac{\mbox{\rule[-5pt]{0pt}{14pt}}\beta_- - 1 - i\varepsilon}
{\mbox{\rule{0pt}{9pt}}\beta_-}\right) + 2\,, \\ [6pt]
B_0(2,4) & = &\;\Delta + \left(1 - \frac{m_Z^2}{u}\right)\ln\left(\frac{
\mbox{\rule[-5pt]{0pt}{14pt}}m_Z^2}{\mbox{\rule{0pt}{9pt}}m_Z^2 - u}\right) +
2\,,
\end{eqnarray}
where $\Delta$ is
\begin{equation}\label{delta}
\Delta = \pi^{(n/2 -2)}\Gamma\left(2 - \frac{n}{2}\right)\,,
\end{equation}
and, again, $\beta_{\pm}$ are given in Eq. (\ref{zroots}).

    The cancellation of the $\ln(m_e^2)$ dependence of Eqs. (\ref{d0z}),
(\ref{c0z1}) and (\ref{c0z2}) can be checked by substituting the explicit
expressions into Eq. (\ref{astu1}).

\section{W Box}

    The diagrams with $W$'s in the loop are the box diagram shown in the second
of Figs.\,1\,(c)  and the triangle diagram given in the last of Figs.\,1\,(c),
together with their counterparts having the Higgs and photon interchanged.
The non-gauge-invariant part of each box diagram is canceled by the
corresponding triangle diagram. The $W$ contribution is determined by two
scalar functions $A_1(s,t,u)$ and $A_2(s,t,u)$, which appear in
Eq.\,(\ref{wbox}). These functions can be related to the $D_{\alpha\beta}$ of
Ref.\cite{pv} as
\begin{eqnarray}
A_1(s,t,u) & = & D_0(s,t,u,m_H^2,m_e^2,m_W^2) +
D_{11}(s,t,u,m_H^2,m_e^2,m_W^2) \nonumber \\
&   & + D_{13}(s,t,u,m_H^2,m_e^2,m_W^2) + D_{25}(s,t,u,m_H^2,m_e^2,m_W^2)\;,\\
A_2(s,t,u) & = & -D_{12}(s,t,u,m_H^2,m_e^2,m_W^2) +
D_{13}(s,t,u,m_H^2,m_e^2,m_W^2) \nonumber \\
&   &+ D_{26}(s,t,u,m_H^2,m_e^2,m_W^2) \;.
\end{eqnarray}
The decomposition into scalar functions takes the form
\begin{eqnarray}\label{a1}
A_1(s,t,u) & = &\;\frac{1}{2}\frac{1}{st}\Biggl[\left(\frac{(s + t - m_W^2)}{t}
\left(m_W^2(t + u) - su\right) - 2m_W^2s\right)D_0(1,2,3,4) \nonumber \\
&  &\;\hspace*{40pt} + (s + t - m_W^2)\Bigl[-\frac{s}{t}C_0(1,2,3) +
\frac{(s^2 + 2st - t^2)}{(s + t)t}C_0(1,2,4) \\
&  &\;\hspace*{40pt} + \frac{(t + u)}{t}C_0(1,3,4) - \frac{u}{t}C_0(2,3,4)
\Bigr] \nonumber \\
&  &\;\hspace*{40pt} + \frac{2s}{(t+u)}\left(B_0(1,3) - B_0(1,4)\right)
+ \frac{2s}{(s + t)}\left(B_0(2,4) - B_0(1,4)\right)\Biggr]\,,
\nonumber\\ [8pt] \label{a2}
A_2(s,t,u) & = &\;\frac{1}{2}\frac{1}{su}\Biggl[\frac{(u - m_W^2)}{u}\left(su +
m_W^2(t + u)\right)D_0(1,2,3,4) \nonumber \\
&  &\;\hspace*{40pt} + (u - m_W^2)\Bigl[\;\frac{s}{u}C_0(1,2,3)
- \frac{(s + t)}{u}C_0(1,2,4)  \\
&  &\;\hspace*{40pt} + \frac{(t + u)}{u}C_0(1,3,4) - C_0(2,3,4)\Bigr] \nonumber
\\
&  &\;\hspace*{40pt} + \frac{2s}{(t + u)}\left(B_0(1,4) - B_0(1,3)\right)
\Biggr]\,. \nonumber
\end{eqnarray}
In this case, $m_1 = m_3 = m_4 = m_W$ and $m_2 = m_{\nu} = 0$ and we follow the
labeling in Fig. 5\,(b).

    The expression for $D_0(1,2,3,4)$ is
\begin{eqnarray}
D_0(1,2,3,4) & = &\;-\frac{1}{su(\lambda_+ - \lambda_-)}\Biggl\{\Bigl[
-Li_2\left(\frac{1 - \lambda_+}{\alpha - \lambda_+ - i\varepsilon}\right) +
Li_2\left(\frac{-\lambda_+}{\alpha - \lambda_+ - i\varepsilon}\right)
\nonumber \\
&  &- Li_2\left(\frac{1 - \lambda_+}{\gamma_+ - \lambda_+ + i\varepsilon}
\right) + Li_2\left(\frac{-\lambda_+}{\gamma_+ - \lambda_+ + i\varepsilon}
\right) \nonumber \\
&  &- Li_2\left(\frac{1 - \lambda_+}{\gamma_- - \lambda_+ - i\varepsilon}
\right) + Li_2\left(\frac{-\lambda_+}{\gamma_- - \lambda_+ - i\varepsilon}
\right) \\
&  &+ Li_2\left(\frac{1 - \lambda_+}{\beta_+ - \lambda_+ + i\varepsilon}
\right) - Li_2\left(\frac{-\lambda_+}{\beta_+ - \lambda_+ + i\varepsilon}
\right) \nonumber \\
&  &+ Li_2\left(\frac{1 - \lambda_+}{\beta_- - \lambda_+ - i\varepsilon}
\right) - Li_2\left(\frac{-\lambda_+}{\beta_- - \lambda_+ - i\varepsilon}
\right)\Bigr]\nonumber \\
&  & -\Bigl[\lambda_+\rightarrow\lambda_-\Bigr]\Biggr\}\,,\nonumber
\end{eqnarray}
and the roots $\alpha$, $\beta_{\pm}$, $\gamma_{\pm}$ and $\lambda_{\pm}$ are
\begin{eqnarray}
\alpha & = &\;1 - \frac{m_W^2}{u}\,,\\
\label{betaw}
\beta_{\pm} & = &\;\frac{1}{2}\left(1\,\pm\,\sqrt{1 - \frac{\displaystyle
\mbox{\rule[-5pt]{0pt}{14pt}}4m_W^2}{\displaystyle\mbox{\rule{0pt}{9pt}}m_H^2}}
\;\right)\,,\\
\gamma_{\pm} & = &\;\frac{1}{2}\left(1\,\pm\,\sqrt{1 - \frac{\displaystyle
\mbox{\rule[-5pt]{0pt}{14pt}}4m_W^2}{\displaystyle\mbox{\rule{0pt}{9pt}}s}}
\;\right)\,,\\
\lambda_{\pm} & = &\;\frac{1}{2}\left(1 + \frac{m_W^2(s - m_H^2)}{-su}\,\pm\,
\sqrt{\left(1 + \frac{\displaystyle\mbox{\rule[-5pt]{0pt}{14pt}}m_W^2(s -
m_H^2)}{\displaystyle\mbox{\rule{0pt}{9pt}}-su}\right)^2 - \frac{\displaystyle
\mbox{\rule[-5pt]{0pt}{14pt}}4m_W^2}
{\displaystyle\mbox{\rule{0pt}{9pt}}s}}\;\right)\,.
\end{eqnarray}

    The $C_0$'s in this case are
\begin{eqnarray}
C_0(1,2,3) & = &\;\frac{1}{s}\Biggl[\frac{\pi^2}{6} - Li_2\left(1 -
\frac{s}{m_W^2}\right) - Li_2\left(\frac{m_W^2}
{m_W^2 - s\gamma_+ -i\varepsilon}\right) \nonumber \\
&  &\;+ Li_2\left(\frac{m_W^2 - s}{m_W^2 - s\gamma_+ -i\varepsilon}\right) -
Li_2\left(\frac{m_W^2}{m_W^2 - s\gamma_- + i\varepsilon}\right) \nonumber \\
&  &\;+ Li_2\left(\frac{m_W^2 - s}{m_W^2 - s\gamma_- + i\varepsilon}\right)
\Biggr]\,, \\[8pt]
C_0(1,3,4) & = &\;\frac{1}{m_H^2 - s}\Biggl[- Li_2\left(\frac{1}{\gamma_+ +
i\varepsilon}\right) - Li_2\left(\frac{1}{\gamma_- - i\varepsilon}\right)
\nonumber \\
&  &\;\hspace*{49pt} + Li_2\left(\frac{1}{\beta_+ + i\varepsilon}\right)
+ Li_2\left(\frac{1}{\beta_- - i\varepsilon}\right)\Biggr]\,, \\[8pt]
C_0(2,3,4) & = &\;\frac{1}{u}Li_2\left(\frac{u}{m_W^2}\right)\,,
\end{eqnarray}
and $C_0(1,2,4)$ being given by Eq. (\ref{c0z3}) with $m_Z\rightarrow m_W$.

    Lastly, the $B_0$'s are
\begin{eqnarray}
B_0(1,3) & = &\;\Delta + \gamma_+\ln\left(\frac{\gamma_+ - 1 + i\varepsilon}
{\gamma_+}\right) + \gamma_-\ln\left(\frac{\gamma_- - 1 - i\varepsilon}
{\gamma_-}\right) + 2\,,\\
B_0(1,4) & = &\;\Delta + \beta_+\ln\left(\frac{\beta_+ - 1 + i\varepsilon}
{\beta_+}\right) + \beta_-\ln\left(\frac{\beta_- - 1 - i\varepsilon}
{\beta_-}\right) + 2\,,\\
B_0(2,4) & = &\;\Delta + \left(1 - \frac{m_W^2}{u}\right)\ln\left(\frac{m_W^2}
{m_W^2 - u}\right) + 2\,,
\end{eqnarray}
where $\beta_{\pm}$ is given by Eq. (\ref{betaw}) and $\Delta$ by Eq.
(\ref{delta}).

\newpage
\begin{figure}[h]
\caption{Typical diagrams for the double pole (a), single pole (c) and box (c)
corrections are shown. An external soild line represents an electron, a wavy
line a gauge boson, a dashed line a Higgs boson and an internal solid line a
fermion, gauge boson, Goldsone boson or ghost.}
\end{figure}

\begin{figure}[h]
\caption{The cross section from the present calculation (solid line) is
compared
with the result from Ref. [2] (dashed line).}
\end{figure}

\begin{figure}[h]
\caption{$\protect\sigma(e\protect\bar{e}\protect\rightarrow \protect\gamma H)$
is plotted for
$m_H$ = 60 (solid), 80 (dashed), 100 (long dashed), 120 (dot dashed),
and 150 (dot dot dashed) GeV.}
\end{figure}

\begin{figure}[h]
\caption{$\protect\sigma(e\protect\bar{e}\protect\rightarrow \protect\gamma H)$
is plotted for
$m_H$ = 60 (solid), 80 (dashed), 100 (long dashed), 120 (dot dashed),
150 (dot dot dashed), 2$m_Z$ (dotted), and 250 (short dashed) GeV.}
\end{figure}

\begin{figure}[h]
\caption{The numbering scheme used for the computation of $D_0(1,2,3,4)$ in the
case of the $Z$ box (a) and the $W$ box (b) are shown.}
\end{figure}

\newpage
\begin{table}[h]
\caption{Contributions to the $Z^0$ double pole corrections.}
\begin{tabular}{c|cc|l|cc}
\multicolumn{6}{c}{${\cal M}_{\mu\nu} = (eg^2/m_Z)(\alpha +
\beta\tan^2\theta_W)
\delta_{\mu\nu}$} \\  \hline
\multicolumn{3}{c}{One-Point} & \multicolumn{3}{c}{Two-Point} \\ \hline
Loop & $\alpha$ & $\beta$ & Loop & $\alpha$ & $\beta$ \\
\tableline
$W$      & $2nA_0$  & $0$     &  $WW$ & $-4(A_0 - m_W^2B_0)$
& $0$  \\
$G$      & $A_0$    & $-A_0$  &  $GG$ & $-(2/n)(A_0 - m_W^2B_0)$
& $(2/n)(A_0 - m_W^2B_0)$ \\
$\theta$ & $-4A_0$  & $0$     &  $\theta\theta$
& $(8/n)(A_0 - m_W^2B_0)$ & $0$
\end{tabular}
\end{table}

\begin{table}[h]
\caption{A comparison of linear and nonlinear gauge contributions to
$\sigma(e\protect\bar{e}\protect\rightarrow H\gamma)$ is presented. For any
contribution, the result of Ref. [1] is given in the left column and that
of the present calculation in the right. The contributions are in femtobarns.}
\begin{tabular}{ccc|cc|cc|cc|cc}
$\sqrt{s}$ &\multicolumn{2}{c}{$|{\cal M}^Z_{\rm Pole}|^2$} &
\multicolumn{2}{c}{$|{\cal M}^{\gamma}_{\rm Pole}|^2$} &
\multicolumn{2}{c}{$2\Re{\it e}({\cal M}^{Z\;*}_{\rm Pole}{\cal M}^{\gamma}_
{\rm Pole})$} & \multicolumn{2}{c}{$|{\cal M}_{\rm Box}|^2$} &
\multicolumn{2}{c}{$2\Re{\it e}({\cal M}^{*}_{\rm Pole}{\cal M}_{\rm Box})$} \\
\tableline
60    &.0069 &.0072 &.012 &.017  &--.0014 &--.0034 &.00027 &.00019 &--.00029 &
--.00036   \\
80    &.21   &.26   &.0038&.024  &--.0044 &--.025  &.0022  &.0015  &.031     &
.025      \\
$m_Z$ &51    &52    &.062 &.066  &.11     &.10     &.0055  &.0039  &.097     &
.026      \\
100   &3.3   &3.2   &.087 &.085  &.083    &.17     &.0084  &.006   &--.29    &
--.25     \\
120   &.61   &.57   &.15  &.14   &.047    &.088    &.025   &.018   &--.28    &
--.24     \\
150   &.53   &.45   &.31  &.26   &.064    &.11     &.14    &.099   &--.74    &
--.57
\end{tabular}
\end{table}
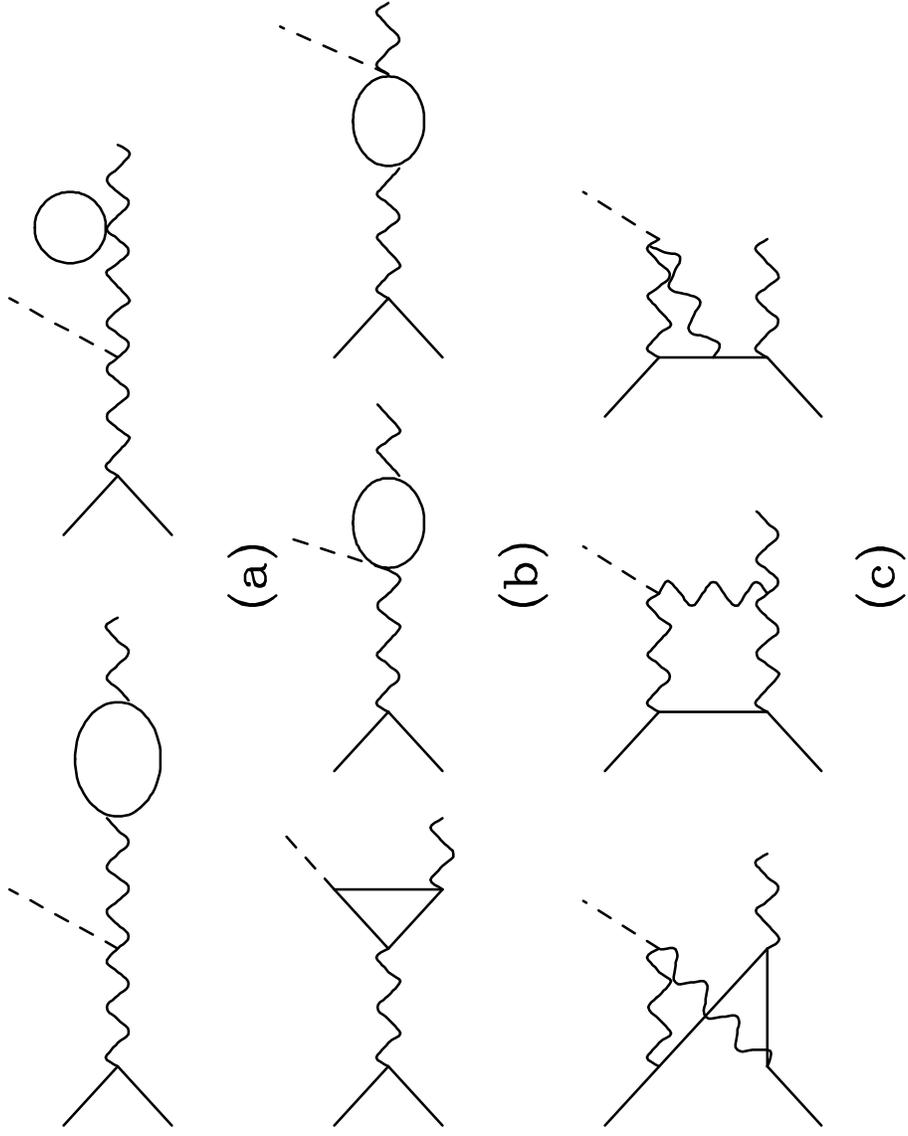
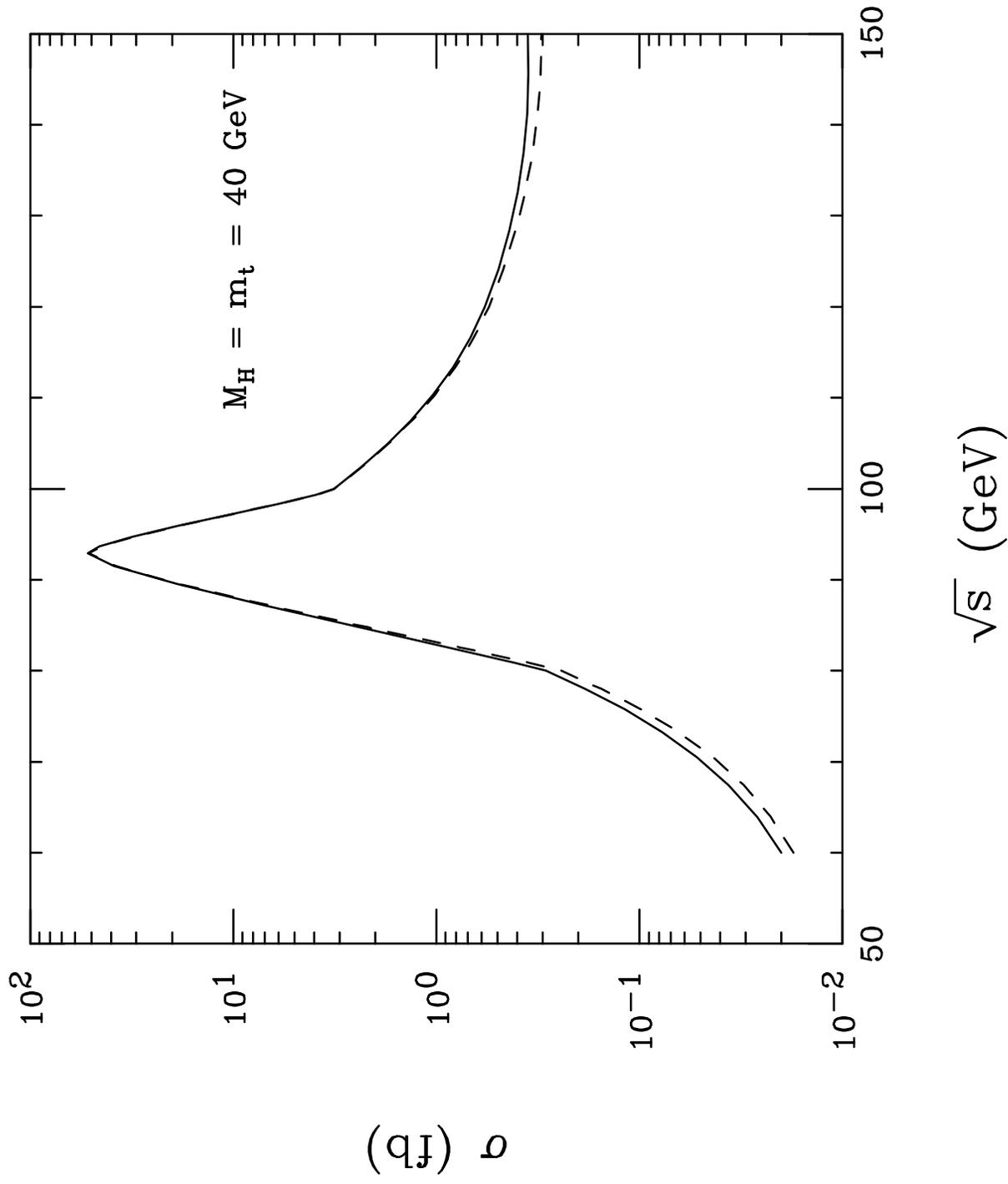
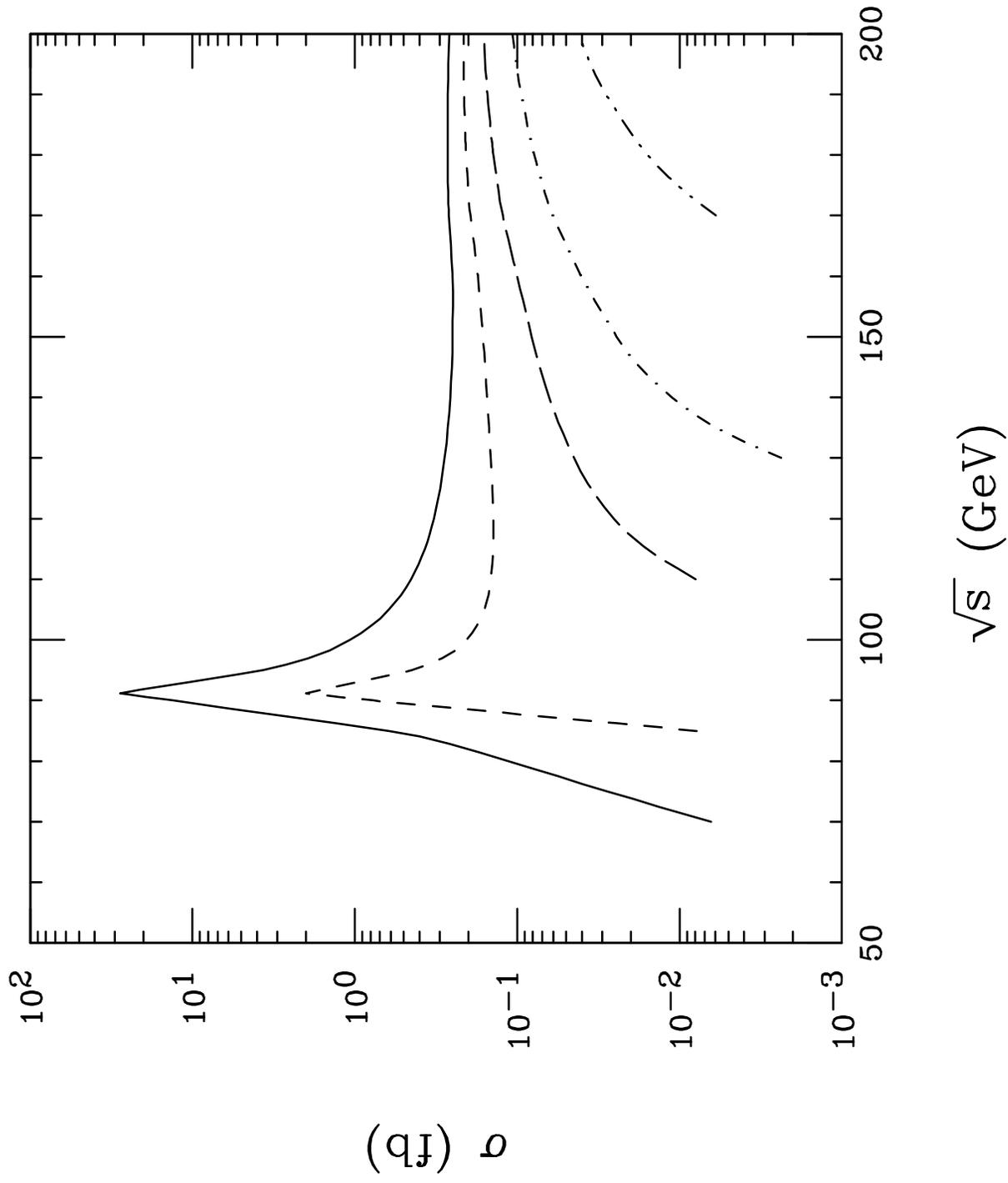
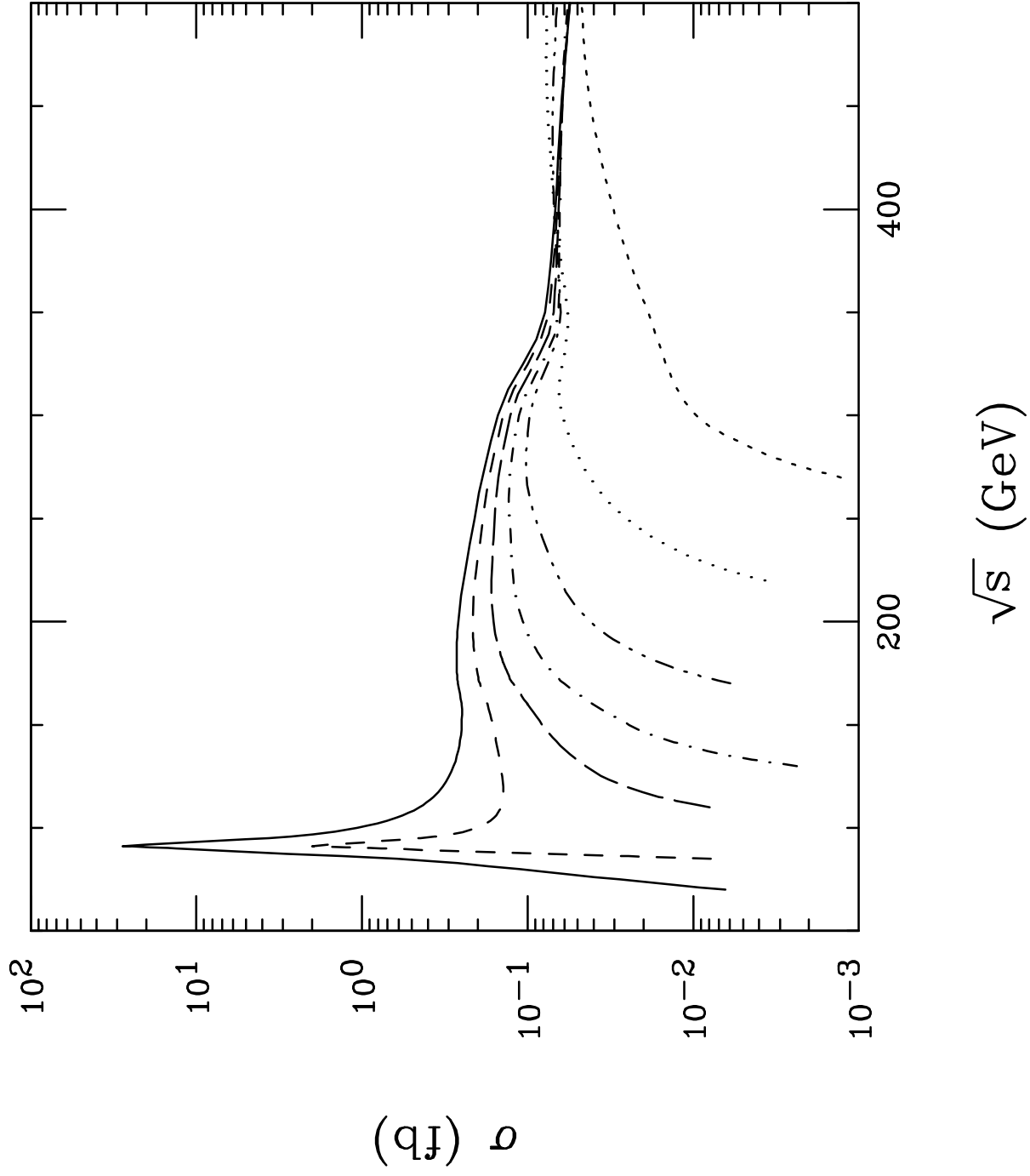
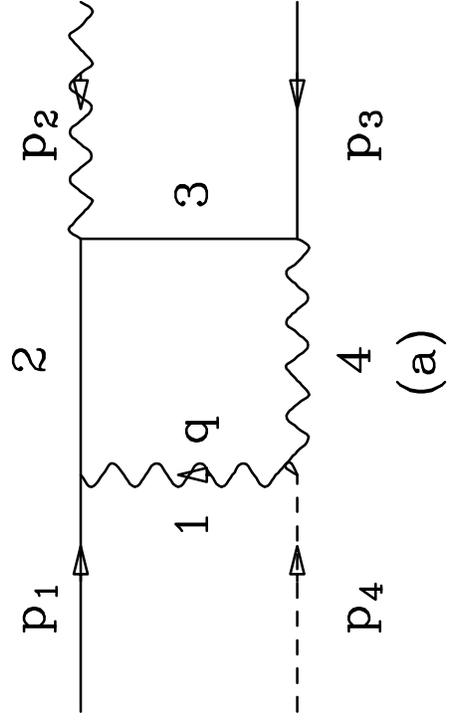
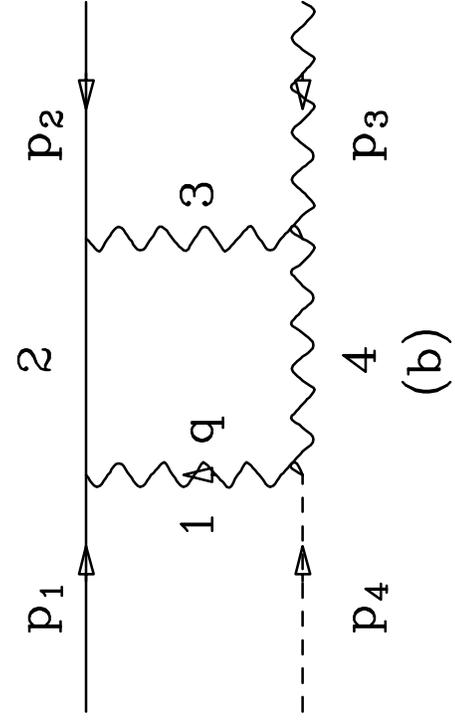
\end{document}